\documentclass[twocolumn,showpacs,amsmath,amssymb,pre,10pt]{revtex4}
\usepackage{amssymb}
\usepackage[final]{graphicx}

\usepackage{epstopdf}

\begin{document}

\title{Matryoshka Locally Resonant Sonic Crystal}
\author{D. P. Elford, L. Chalmers, F. Kusmartsev and G. M. Swallowe}

\affiliation{Department of Physics, Loughborough University,
Loughborough, LE11 3TU, United Kingdom\\}

\begin{abstract}
The results of numerical modelling of sonic crystals with resonant array elements are reported. The investigated resonant elements include plain slotted cylinders as well as various their combinations, in particular, Russian doll or Matryoshka configurations. The acoustic band structure and transmission characteristics of such systems have been computed with the use of finite element methods. The general concept of a locally resonant sonic crystal is proposed, which utilises acoustic resonances to form additional band gaps that are decoupled from Bragg gaps. An existence of a separate attenuation mechanism associated with the resonant elements, which increases performance in the lower frequency regime has been identified. The results show a formation of broad band gaps positioned significantly below the first Bragg frequency. For low frequency broadband attenuation a most optimal configuration  is the Matryoshka sonic crystal, where each scattering unit is composed of multiple concentric slotted cylinders. This system forms numerous gaps in the lower frequency regime, below Bragg bands, whilst maintaining a reduced crystal size viable for noise barrier technology. The finding opens new perspectives for construction of sound barriers in the low frequency range usually inaccessible by traditional means including conventional sonic crystals.
\end{abstract}


\maketitle

Recent years have seen a growing interest in the potential for the use of sonic crystals as noise barriers, with reported sound attenuation up to 20 dB \cite{Perez: Sound} and 25 dB \cite{Sanchez:Acoustic}. Such crystals usually consist of periodic arrays of a high mechanical impedance material (often as cylindrical rods) and are known to give high attenuation at selective but often rather narrow frequency bands as a consequence of multiple scattering phenomena. An advantage of sonic crystals noise barriers is that, by varying the distance between the scatterers, it is possible to attain peaks of attenuation in a selected frequency range. Further advantages of a sonic crystal barrier in comparison with more traditional solid sound barriers, are its ability to allow light to pass and, uniquely, that it does not present an obstruction to the free flow of air. The relationship between the lattice parameter and operating frequency suggest extremely large barriers will be required to attenuate lower frequency noise such as traffic. Therefore locally resonant sonic materials (LRSM) \cite{Liu: Locally} are better suited due to their ability to form band gaps decoupled from the periodicity. However, these band gaps cover a narrow attenuation range and such LRSM are unsuitable for use as a noise barrier.

We investigate the effects of elastic wave propagation through a new class of LRSM with multiple acoustic resonances, capable of broadening the range of attenuation. The proposed sonic crystal forms broad attenuation bands in the lower frequency regime and comprises concentric slotted cylinders. The preliminary results of this work is presented in Refs \cite{chalmers}. Previously Hu \emph{et al.} \cite{Hu: Two} constructed a sonic crystal lens composed of an array of two-dimensional Helmholtz resonators, which in the long-wave regime was found to have a high relative acoustic refractive index $n$ and at the same time, a small acoustic impedance $Z$ mismatch with air for airborne sound. Movchan \emph{et al.} \cite{movchan} investigated the asymptotic analysis of an Eigenvalue problem for the Helmholtz operator in a periodic structure involving split-ring resonators. Furthermore, the wave propagation in a sonic crystal with Helmholtz resonator defect was studied by Wu \emph{et al.} \cite{Wu}, where a Helmholtz resonator is placed as a point defect of the sonic crystal and exhibits local resonance phenomena. In the present paper an array of the resonant elements that have resonances below the Bragg band gaps have been studied. In particular the elements having a shape of slotted cylinders and their various configurations have been considered. The interaction between their resonances produces band gaps and  gives rise to a new form of acoustic attenuation. The proposed systems have been studied numerically with the use of finite elements methods (FEM). So far the results are in complete agreement with laboratory experiments, see also Refs \cite{chalmers}.

\section{Numerical modelling of resonant arrays with the use of Finite Element Method}
\subsection{Eigenvalue Analysis}
\begin{figure}[htp!]
		\includegraphics[angle=0,width=0.5\columnwidth]{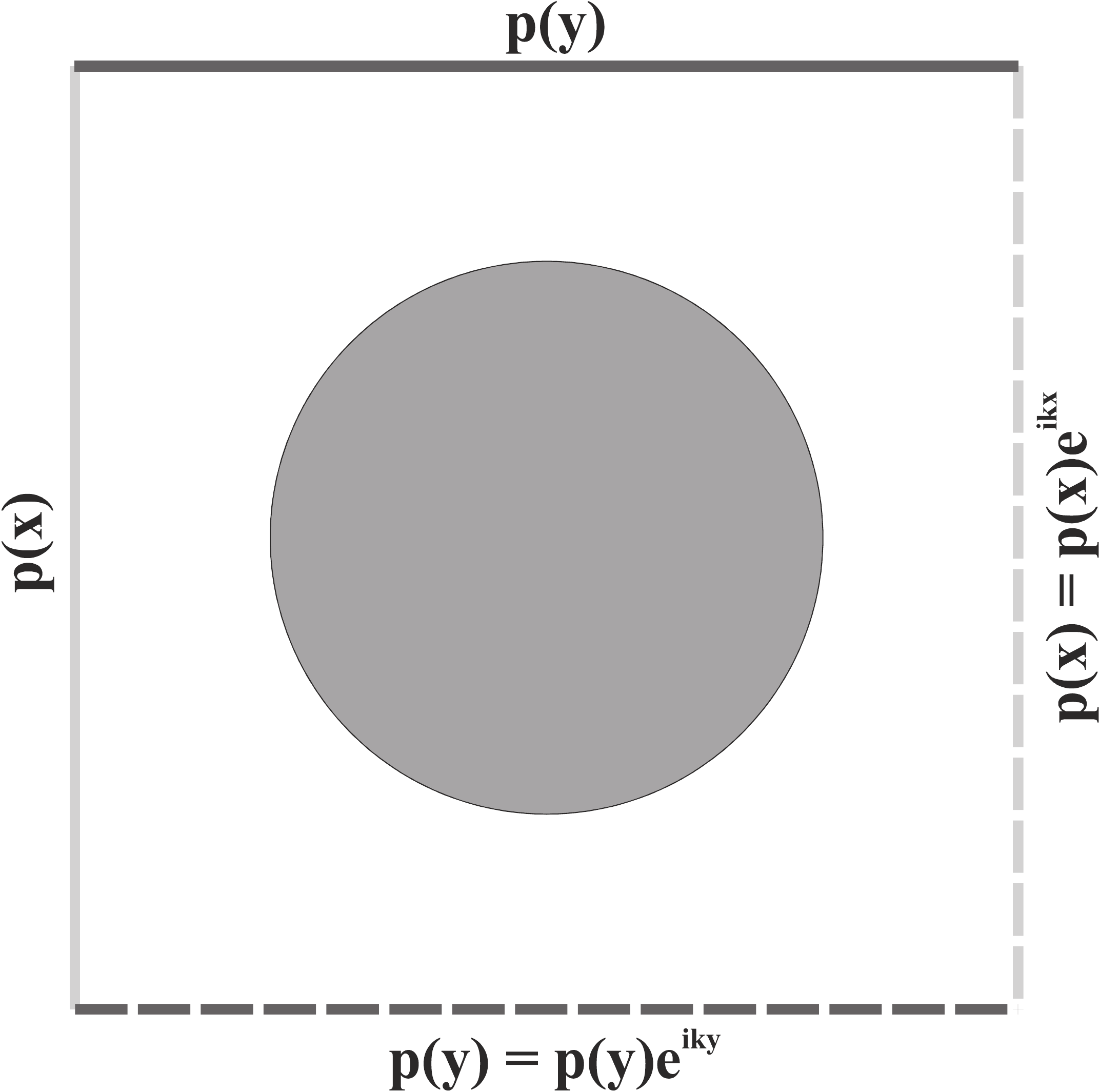}
		\vskip -0.5mm \caption{Single unit cell of an infinite sonic crystal system with periodic boundary conditions applied to the borders of the unit cell.}
	\label{unitcell}
\end{figure}

\begin{figure}[htp!]
	\centering
		\includegraphics[angle=0,width=0.85\columnwidth]{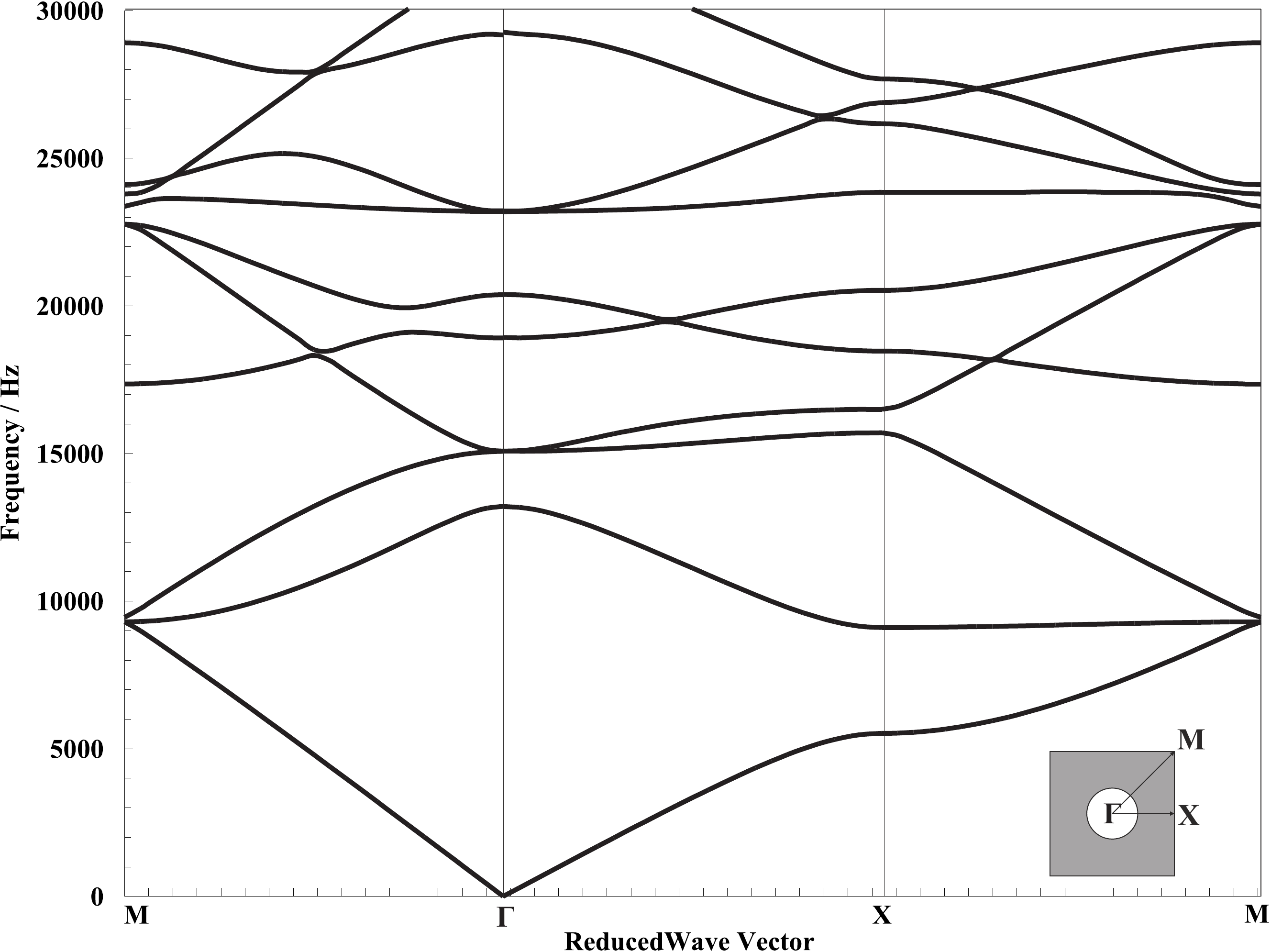}
		\vskip -0.5mm \caption{Finite Element computed band structure for a sonic crystal consisting of steel scatterers embedded in air ($r$ = 6.5 mm, $a$ = 22 mm). Inset: Brillouin zone. $\Gamma X$ refers to the [1 0] direction, and $\Gamma M$ the [1 1] direction, while $XM$ refers to the wavevector varying from [1 0] to [1 1] on the side of the Brillouin zone.}
	\label{comsolsteelbraggband}
\end{figure}

To verify methods used first we consider infinite arrays of solid cylinders. Their band structures are obtained using the FEM which  was developed in the framework of Comsol Multiphysics \cite{Comsol}. For a sonic crystal in a two-dimensional square array, the unit cell (seen in Figure \ref{unitcell}) is used as a basis for the calculations. 
The structure is assumed to be infinite and periodic in the  direction $x$  with the period $a_1$ and in the direction $y$ with the period $a_2$ and described by two basis vectors: $(a_1,0)$ and $(0,a_2)$. According to the Floquet-Bloch theorem, the relation for the pressure distribution $p$ for nodes lying on the boundary of the unit cell can be expressed as:

\begin{equation}
p(\vec{x}+\vec{a}_1+\vec{a}_2) = p(x) \exp[i(k_x a_1+k_y a_2)],
\end{equation}

where $x$ is the position vector in the unit cell and $\vec{k}=(k_x,k_y)$ is the Bloch wavevector.
Considering the periodic boundary conditions above allows the reduction of the model to a single unit cell. First we apply boundary conditions of the Neumann type, this is required on boundaries where pressure $p$ is controlled by a periodic boundary condition. Next a phase relation is applied in the boundary of the unit to define boundary conditions between adjacent units. This phase relation is related to the wavenumber of the incident wave in the periodic structure. The periodic boundary conditions are applied to truncate the two-dimensional simulation plane in the $x$ and $y$ directions, by reducing the system to one unit cell. An ideal crystal is infinitely periodic, hence the periodic boundary condition ensures that the finite simulation space mimics an infinitely periodic crystal in the $x$ and $y$ directions. The pressure components at all edges of the computational domain are relocated by the periodic boundary conditions to the opposite edges of the domain. This enforces the condition that a wave travelling into the top edge of the computational domain is relocated and appears outside the computational domain in the bottom periodic boundary condition. Similarly a wave travelling into the bottom edge of the domain is relocated and appears outside the computational domain in the top periodic boundary condition domain. Similarly this occurs for the left and right edges of the domain. The cylinder in the unit cell is considered to be rigid, and therefore the Neumann boundary condition is applied to its surface.
By defining the Bloch wavevector in the first Brillouin zone, for the $\Gamma X$ direction, $k_x$ varied from 0 to $\pi$, whilst $k_y = 0$; $\Gamma M$ direction $k_y$ varied from 0 to $\pi$, whilst $k_x = \pi$; and in the $XM$ direction $k_x$ and $k_y$ varied from 0 to $\pi$. The analysis of the first ten Eigenfrequencies and the corresponding Eigenvectors is computed. The Eigenvectors are related to the pressure distribution of the mode. In this investigation the infinite sonic crystal is composed of steel cylinders in air, with lattice parameter $a$ = 22 mm, radius of steel scatterer $r$ = 6.5 mm, and packing fraction $f$ = 0.27. Figure \ref{comsolsteelbraggband} displays the characteristic band structure for this system and is plotted in the three principal symmetry directions.

The dispersion remains isotropic in the low-frequency range, following a linear trend $c = \frac{\omega}{k}$, where the propagating wave cannot resolve the fine structure of the cylinders in the long-wavelength limit. A sonic band gap opens between the first two bands in the $\Gamma X$ direction. It can be seen that toward the edges of the Brillouin zone, the dispersion is no longer linear, with a curving of the bands where, at the edge, the bands exhibit zero group velocity.

\subsection{Transmission Analysis}
The finite element method has been utilised to calculate the pressure field behind a sonic crystal and to generate a pressure map of the system at fixed frequencies. The Comsol Multiphysics software is adopted to solve the acoustic wave propagation in the sonic crystals. The equation used to analyse the acoustic wave problems is expressed as

\begin{equation}
\frac{1}{\rho_{0} c^2} \frac{\partial^2p}{\partial t^2}+\nabla \cdot\left(-\frac{1}{\rho_{0}}\nabla p\right)=0.
\end{equation}	
	
This reduces to a Helmholtz equation for a time harmonic pressure wave excitation, $p = p_0e^{i\omega t}$
\begin{equation}
\nabla \cdot \left(-\frac{1}{\rho_{0}}\nabla p_0\right)-\frac{\omega^2 p_{0}}{\rho_{0} c^2}=0,
\label{helmholtzeq}
\end{equation}

where $\omega = 2\pi f$ is the angular frequency. By solving equation (\ref{helmholtzeq}), the pressure field can be obtained.

A two-dimensional sonic crystal system in a square lattice is described in Comsol Multiphysics, with lattice parameter $a$ = 22 mm, cylinder radius $r$ = 6.5 mm has been calculated. Material parameters for this system are $\rho_s$ = 7800 kgm$^{-3}$, $c_s$ = 6100 ms$^{-1}$, $\rho_a$ = 1.25 kgm$^{-3}$, $c_a$ = 343 ms$^{-1}$. In the case of the rigid cylinders in the sonic crystal system, sound-hard boundary conditions have been applied; i.e. the normal component of the velocity of the air particles is zero in the walls of the cylinders. The radiation boundary conditions at the exterior edges of the rectangular domain are considered to be perfectly absorbing. In the simulations, a rising tone noise source at the left edge of the domain, from 0 -- 30000 Hz, is modelled as a radiation condition with pressure source set to 1 Pa, which is equivalent to a 90 dB source. For the numerical simulation, we use a triangular mesh of approximately $10^6$ elements to solve the wave equation across the domain. 
\begin{figure}[htp!]
\includegraphics[width=0.85\columnwidth]{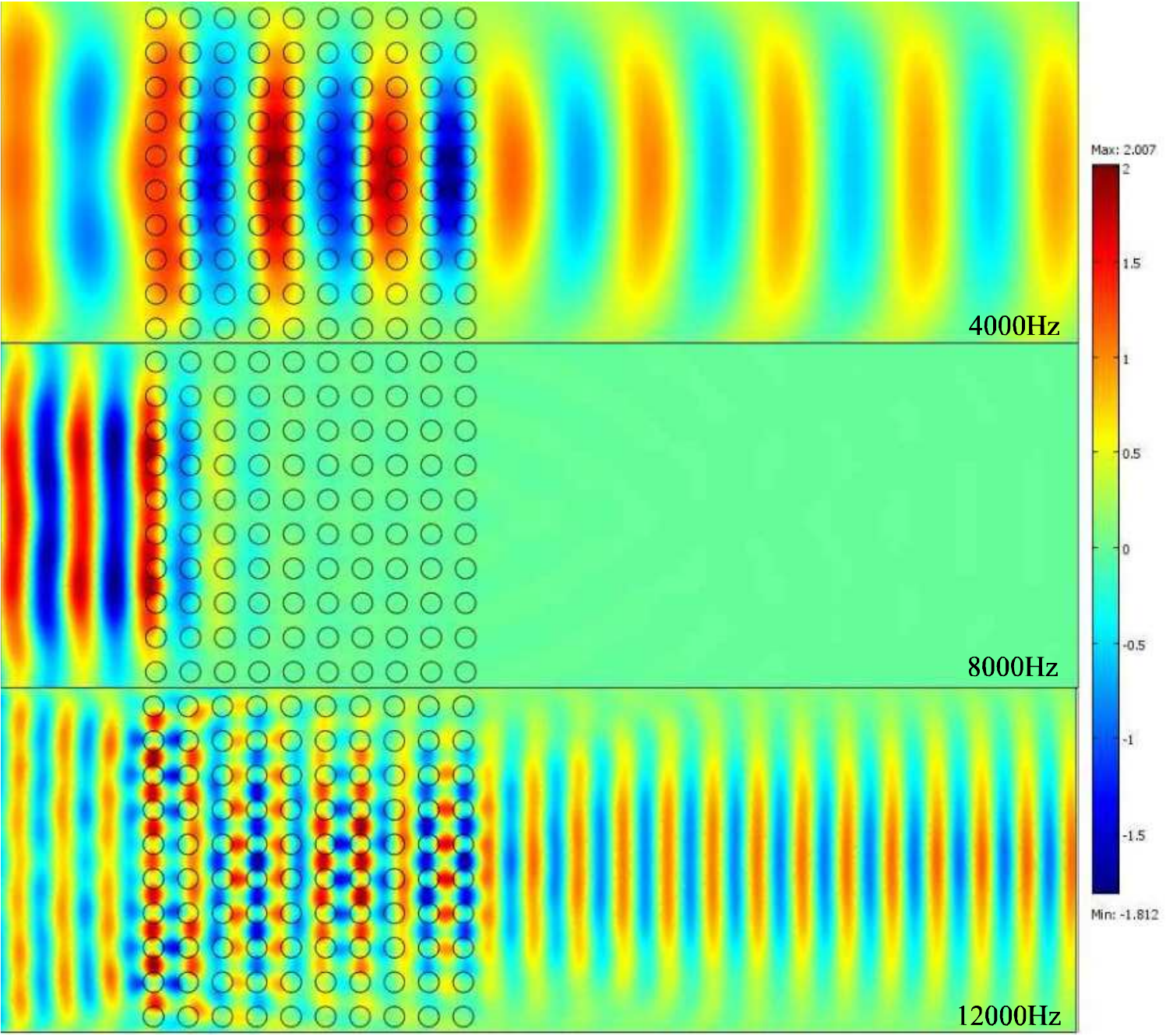}
\vskip -0.5mm \caption{(Colour online) Finite Element computed pressure maps for solid steel cylinders in air taken at three frequencies: 4000, 8000 and 12000 Hz.}
\label{steelbraggplots}
\end{figure}
The illustrated pressure maps are taken at 4000 Hz (pre Bragg band gap formation), 8000 Hz (in the centre of the band gap) and 12000 Hz (after Bragg band gap formation) as shown in Figure \ref{steelbraggplots}. The pressure map taken pre band gap formation clearly demonstrates that at low frequencies the sonic crystal system behaves as a homogeneous material and acoustic wave propagation is unaffected by the periodic structure. This is due to the lattice parameter being much smaller than the relevant wavelength. The pressure map at 8000 Hz, in the centre of the band gap clearly shows band gap formation, with the wavelength of the incoming acoustic wave comparable to the lattice parameter. The acoustic wave is severely attenuated due to multiple scattering effects and a shadow zone is formed behind the sonic crystal. At 12000 Hz, post band gap formation, the wavelength of the acoustic wave is smaller than that of the lattice parameter of the sonic crystal system. The wave is free to propagate through the sonic crystal system, as the plane wave cannot resolve the individual scatterers. 

\begin{figure}[htp!]
\centering
\includegraphics[width=0.85\columnwidth]{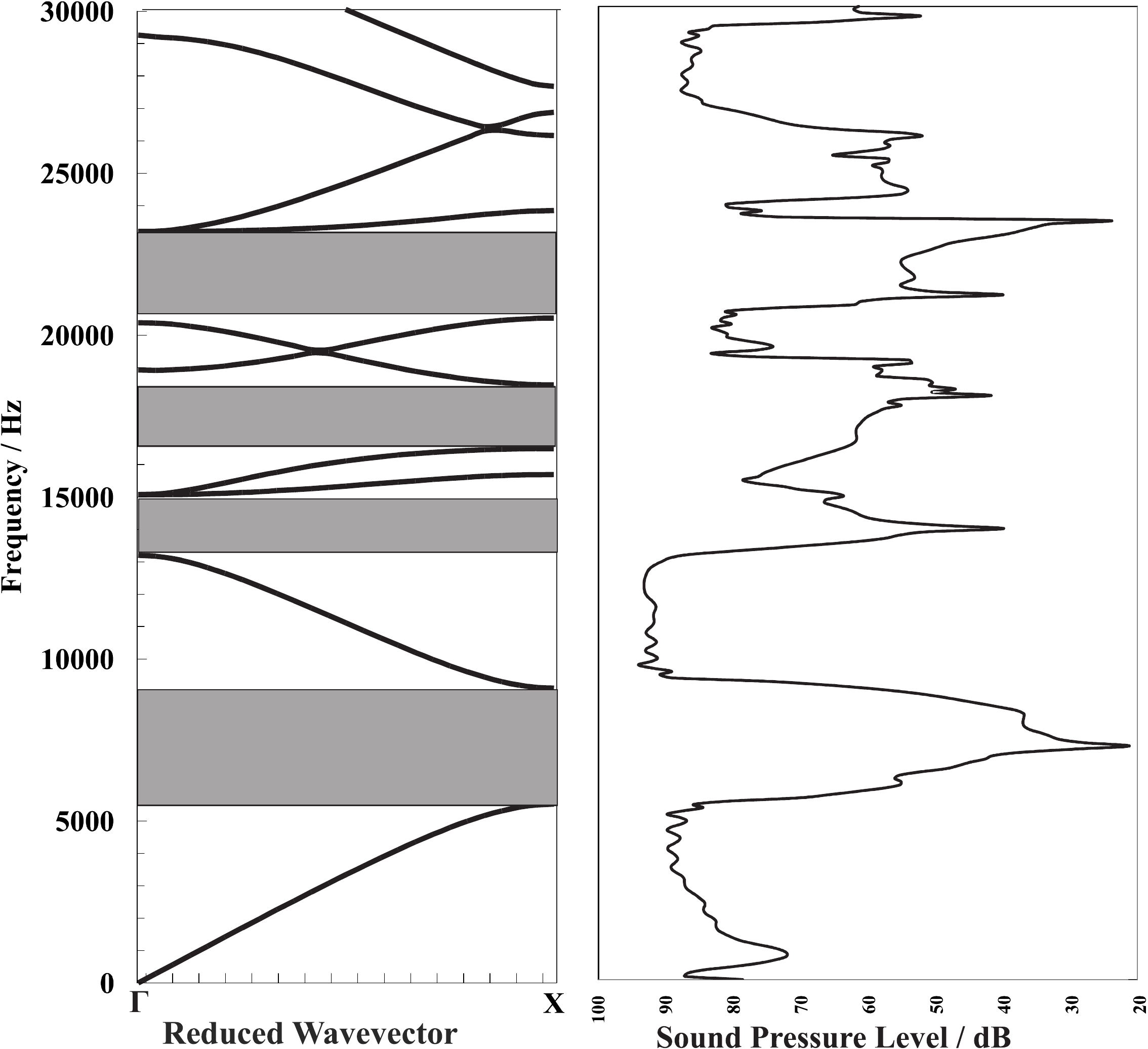}
\vskip -0.5mm \caption{A comparison of Finite Element computed band structure in the $\Gamma X$ direction against the Finite Element computed frequency spectra for sound pressure level for conventional sonic crystal.}
\label{band30000}
\end{figure}

By solving for a parametric sweep of frequency, a frequency spectrum displaying the attenuation properties of the sonic crystal can be constructed. A comparison of the transmission spectrum from 0 -- 30000 Hz against the computed band structure, limiting the study to the $\Gamma X$ direction can be seen in Figure \ref{band30000}. The finite element transmission calculations give band gaps of larger width than those predicted by the band structure. This difference could be attributed to the finite number of scatterers used in the transmission simulations and the subsequent diffraction effects around the edges of the sonic crystal structure. The overall position of the band gaps are in good agreement with those of the band structure calculations.

\section{C-shaped Locally Resonant Sonic Crystal}
\begin{figure}[htp!]
	\centering
		\includegraphics[width=0.5\columnwidth]{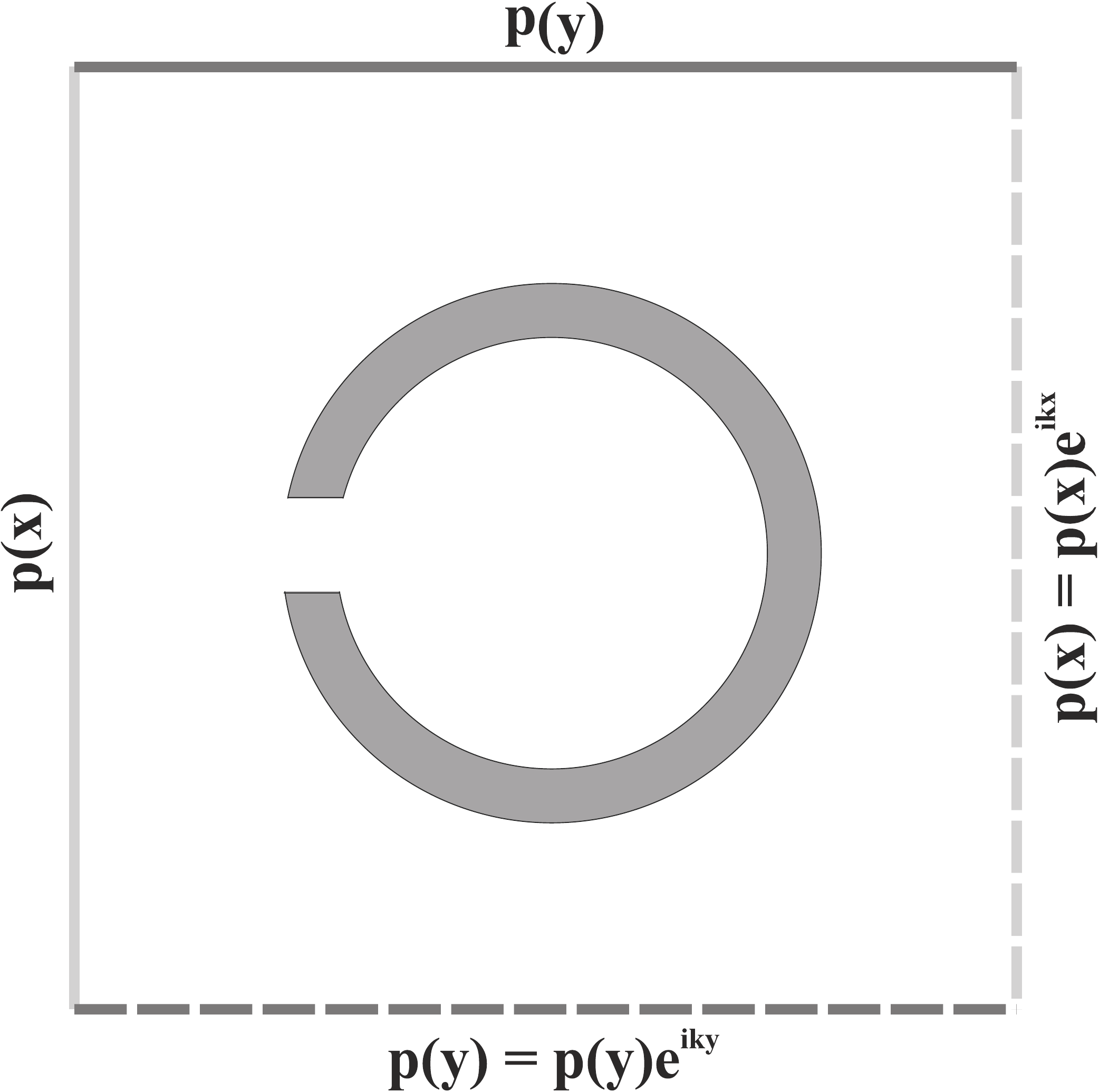}
		\vskip -0.5mm \caption{Unit cell for a C-shaped locally resonant sonic crystal with periodic boundary conditions described.}
\label{unitcellres}
	\end{figure}

The conventional sonic crystal modelled with solid scattering inclusions form band gaps solely due to the periodicity in agreement with theory and our experiments \cite{chalmers}. To operate below this Bragg gap we now investigate a design of locally resonant sonic crystal (LRSC) which is an array of slotted cylinders.
An advantage of using Comsol Multiphysics to compute acoustic band structure is the capability of modelling more complex scatterer geometries. Similar to the conventional sonic crystal system modelled previously, periodic boundary conditions have been employed, see Figure \ref{unitcellres}. 
	
\begin{figure}[htp!]
	\centering
		\includegraphics[angle=0,width=0.85\columnwidth]{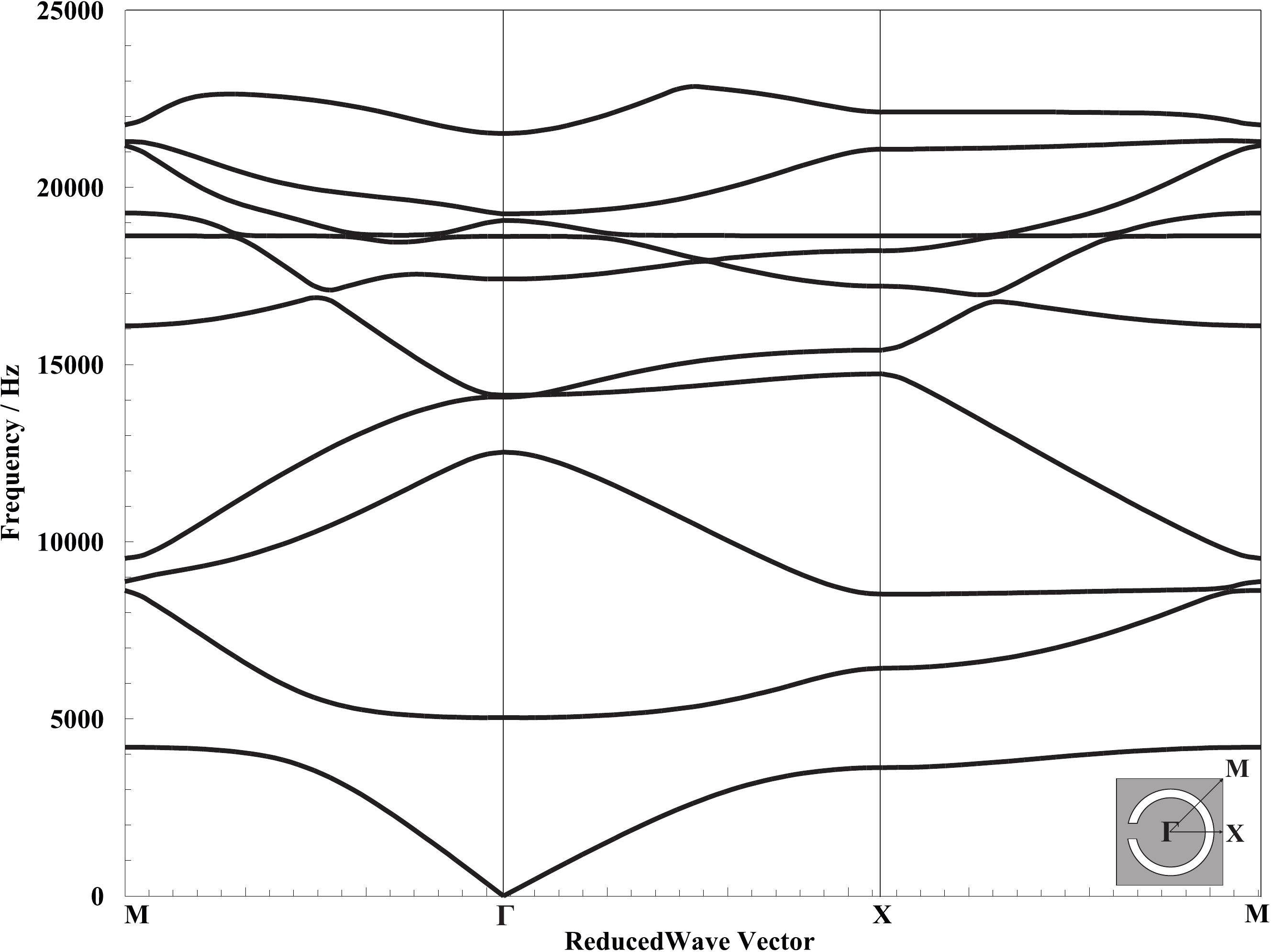}
		\vskip -0.5mm \caption{Finite Element computed band structure for a C-shaped locally resonant sonic crystal.}
	\label{resband}
\end{figure}

Again, by varying the wavevector in the first Brillouin zone for the first ten Eigenvalues, the band structure can be constructed, see Figure \ref{resband}. The figure gives the computed band structure of a two-dimensional sonic crystal, comprising slotted tubes with inner radius 5 mm, external radius 6.5 mm and slot width 4 mm arranged in a square lattice in air. The period is 22 mm. We call each resonating inclusion, a C-shaped resonator.

We note the appearance of a flat band in the band structure (Figure \ref{resband}). Modes associated with a flat band should have a group velocity equal to zero and exhibit strong spatial localisation. In practice, such localised modes are often created by inserting a defect in a periodic structure, i.e. creating a cavity \cite{Wu}. It is clear that the acoustic resonance owing to the C-shaped inclusions leads to the appearance of this flat band, forming a complete acoustic band gap that is induced by the local acoustic resonance of each individual scatterer. The slotted tubes act analogous to Helmholtz resonators and all have the same resonance frequency, $f_{res}$ = 4840 Hz. The combined action of the resonators induces the degenerate state to form a band gap symmetrically around $f_{res}$ spanning 4190 -- 5190 Hz, centred at 4690 Hz. 

Due to the periodicity of the C-shaped LRSC, this structure still exhibits Bragg band gaps, the first of which spans 6410 -- 8550 Hz. A further three Bragg bands are present due to the fulfillment of the Bragg condition located at 12525 -- 14135 Hz, 15400 -- 17210 Hz and 21050 -- 22140 Hz. The introduction of the extra, flat resonance band could lead to the construction of viable acoustic barriers in the low frequency regime, which offer sound attenuation in all crystal lattice planes. The flat band (originating from the localised acoustic resonance seen in the band structure) is a large anticrossing gap; this is generally referred to as a hybridization gap in the context of sonic crystals \cite{Sainidou}. 

\begin{figure}[htp!]
	\centering
		\includegraphics[trim = 0mm 0mm 0mm 0mm,clip,angle=0,width=0.85\columnwidth]{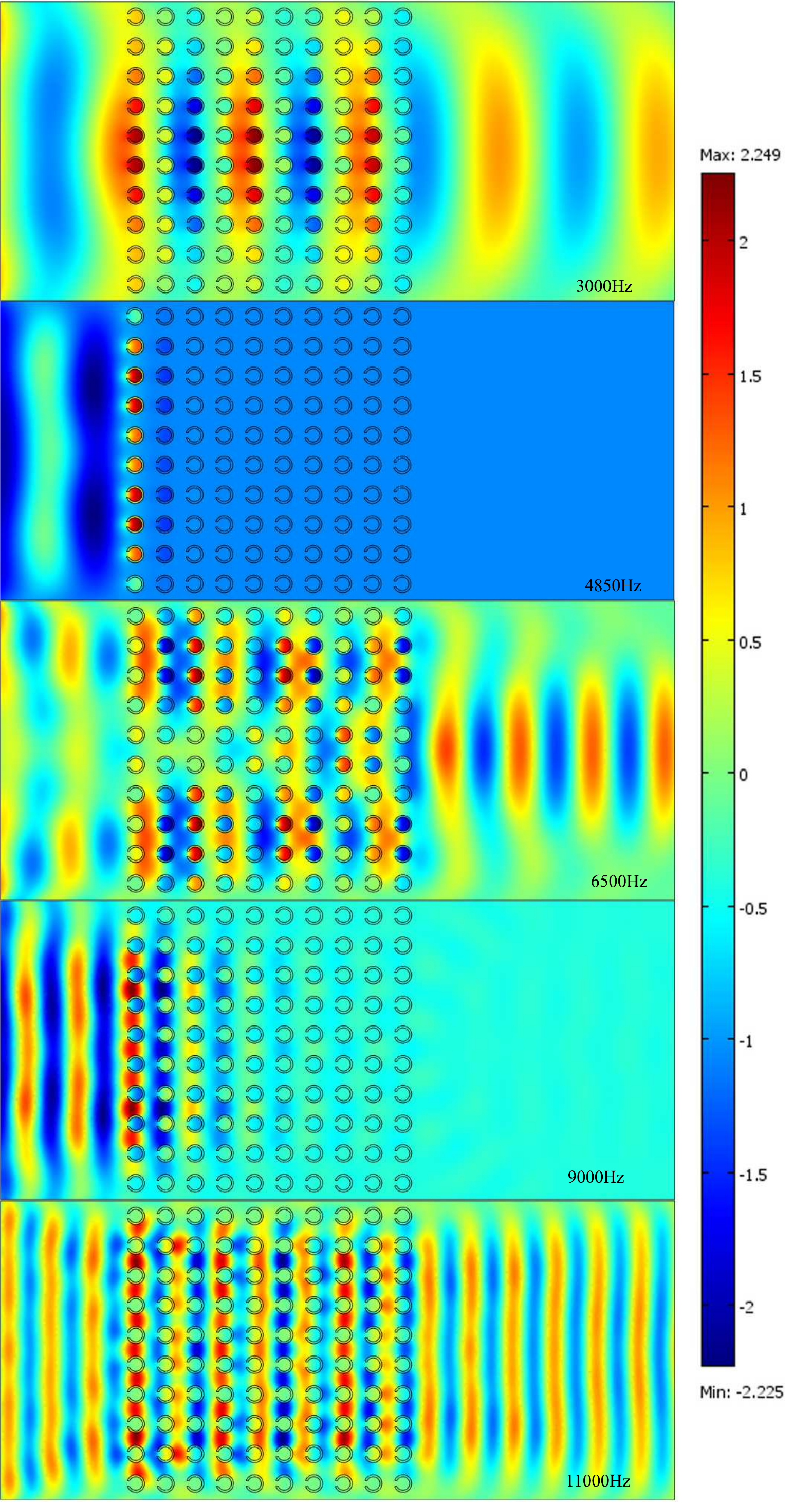}
		\vskip -0.5mm \caption{(Colour online) Finite Element computed pressure level maps for a C-shaped locally resonant sonic crystal at 3000, 4850, 6500, 9000 and 11000 Hz.}
\label{steelrespmap}
	\end{figure}
	
FE transmission simulations are implemented for the C-shaped LRSC. Similar boundary conditions have been applied as for the conventional sonic crystal investigation. Effectively this new system is a duplicate of the conventional sonic crystal system detailed above, but with the inclusion of a slot to create a resonant cavity.

Computed pressure maps, taken at five frequencies of interest, demonstrate the propagation of an acoustic plane wave through the C-shaped LRSC. Similar to the conventional sonic crystal, at frequencies below the active frequency (3000 Hz) of the sonic crystal, the incoming wave propagates as if the system was a homogeneous medium. The computed pressure map taken at 4850 Hz clearly shows that the C-shaped LRSC attenuates the wave in this region. The pressure map, see Figure \ref{steelrespmap}, indicates that regions of maximum pressure are localised to the inclusions, at the resonance frequency. In the higher frequency region, around 6500 Hz, the acoustic wave is free to propagate through the system. As we approach the Bragg band gap frequency of 9000 Hz, it is clear again to see the appearance of band gap type attenuation. Looking at the pressure maps it becomes apparent that the two regions of attenuation appear to be controlled by two different mechanisms.	
	
\begin{figure}[htp!]
	\centering
		\includegraphics[trim = 0mm 0mm 0mm 0mm, angle=0,clip,width=0.85\columnwidth]{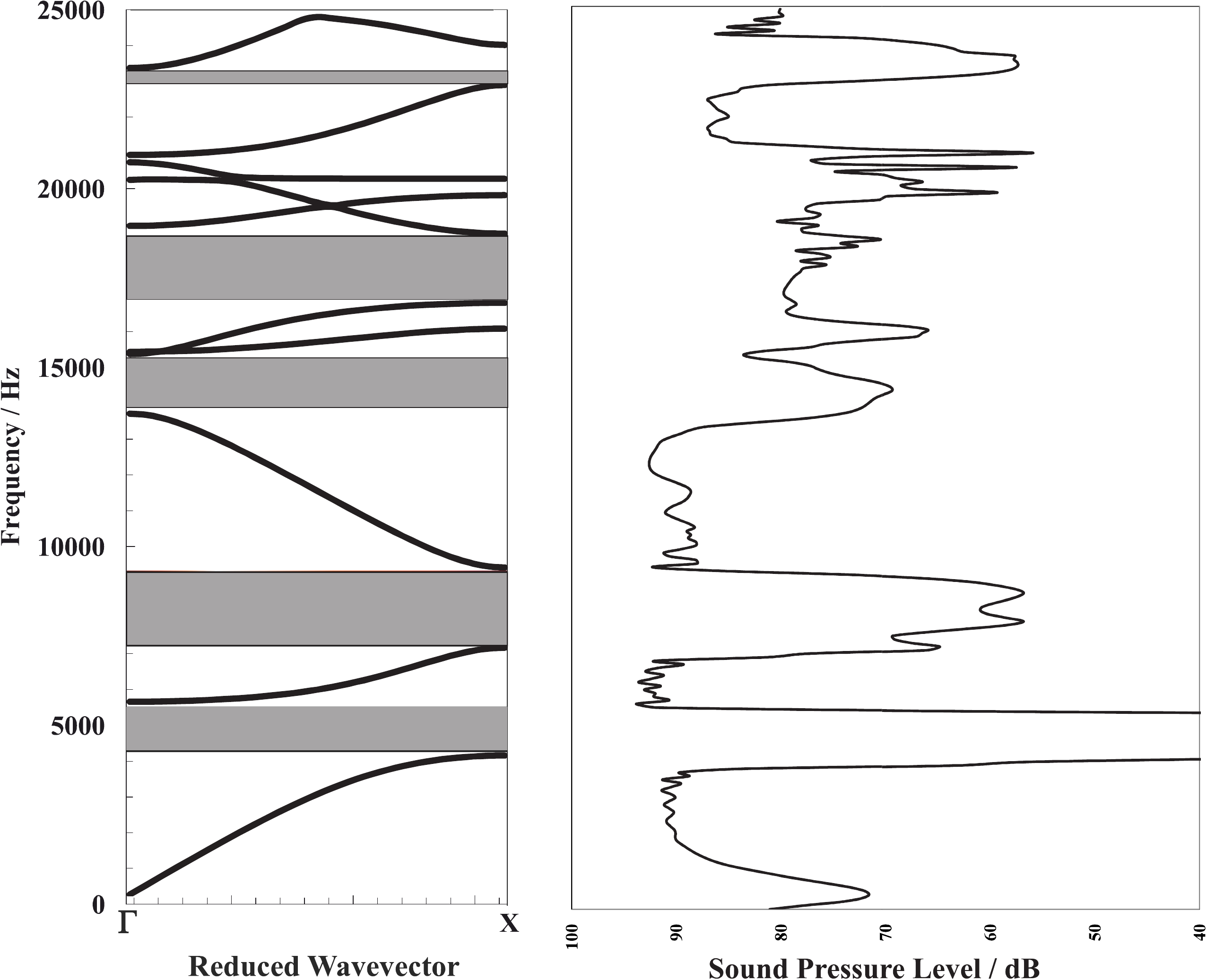}
		\vskip -0.5mm \caption{A comparison between Finite Element computed band structure and Finite Element transmission simulation for a C-shaped locally resonant sonic crystal. }
	\label{res25000}
\end{figure}

In the transmission spectrum, see Figure \ref{res25000}, additional attenuation peaks can be observed ($\sim$20000 Hz). If we compare the location of these peaks with the computed band structure, the peaks appear to be attributed to anticrossing regions present in the band structure. In general, such gaps originate from level repulsion, when two bands of the same symmetry avoid crossing each other. The appearance of these anticrossing regions are beyond the scope of this investigation, but should be investigated further to enhance the performance of the C-shaped LRSC. The reader is directed toward a seminal paper by Wu \emph{et al.} \cite{wu:level} detailing this phenomenological effect. The physical origin of these anticrossing gaps, is different when compared with those induced by the acoustic resonance. The flat band (hybridisation gap) originating from the acoustic resonance of the C-shape scatterer and regions corresponding to the anticrossing gaps that are formed due to the longitudinal displacement field in the homogeneous effective medium \cite{Sainidou}. The narrowness of the anticrossing gaps indicate that they are much weaker. The hybridization discussed is analogous to s-d hybridization in the energy band structure of transition metals, see for example Harrison \cite{harrison}.

\section{Matryoshka Sonic Crystal}
\begin{figure}[!htp]
	\centering
		\includegraphics[trim = 0mm 0mm 0mm 0mm,clip,angle=0,width=0.5\columnwidth]{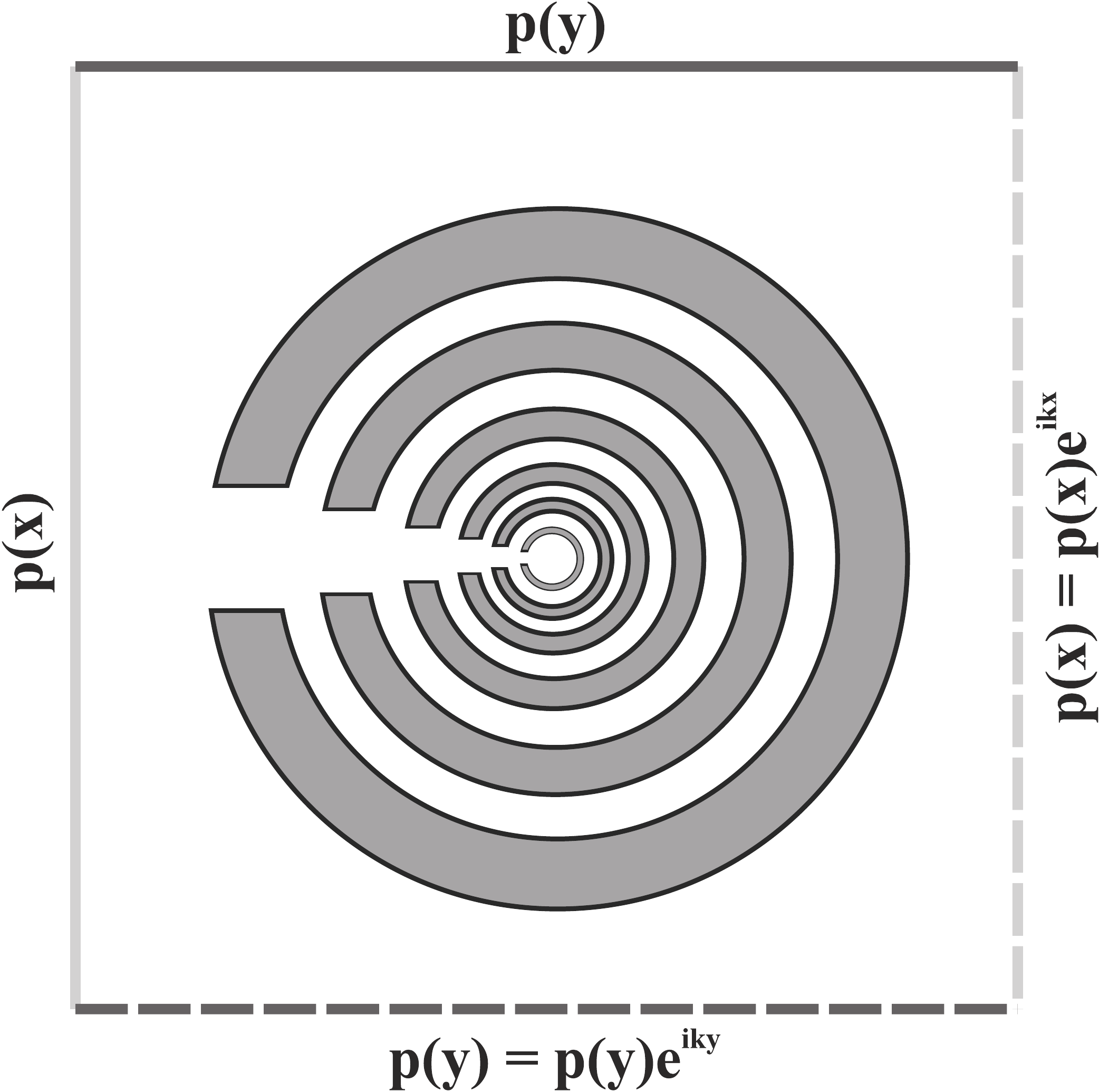}
		\vskip -0.5mm \caption{Schematic of unit cell used in band structure calculations for a six concentric Matryoshka system.}
\label{6matunit}
	\end{figure}

The C-shaped tubes act as acoustic resonators which give rise to a single flat band that extends across all high symmetry directions and is located below the Bragg gap. Its position is dependent upon the cavity dimensions and is independent of the sonic crystal periodicity. For practical applications of sonic crystals as noise barriers it is desirable to be able to broaden the width of this resonance gap. One method to achieve this is to include multiple resonator sizes and `overlap' the individual resonance peaks. We have investigated mixed arrays that display this ability \cite{chalmers}, however, in order to save space and reduce the overall barrier thickness we know propose a design of sonic crystal with resonators placed concentrically inside one another. We coin this the Matryoshka (Russian doll) configuration. Specifically we investigate a Matryoshka sonic crystal whose unit cell is defined with six concentric C-shaped resonators, all tuned to frequencies that lie within 200 Hz of each other in the low frequency regime, (see Figure \ref{6matunit}).

\begin{figure}[!htp]
	\centering
		\includegraphics[trim = 0mm 0mm 0mm 0mm,clip,angle=0,width=0.85\columnwidth]{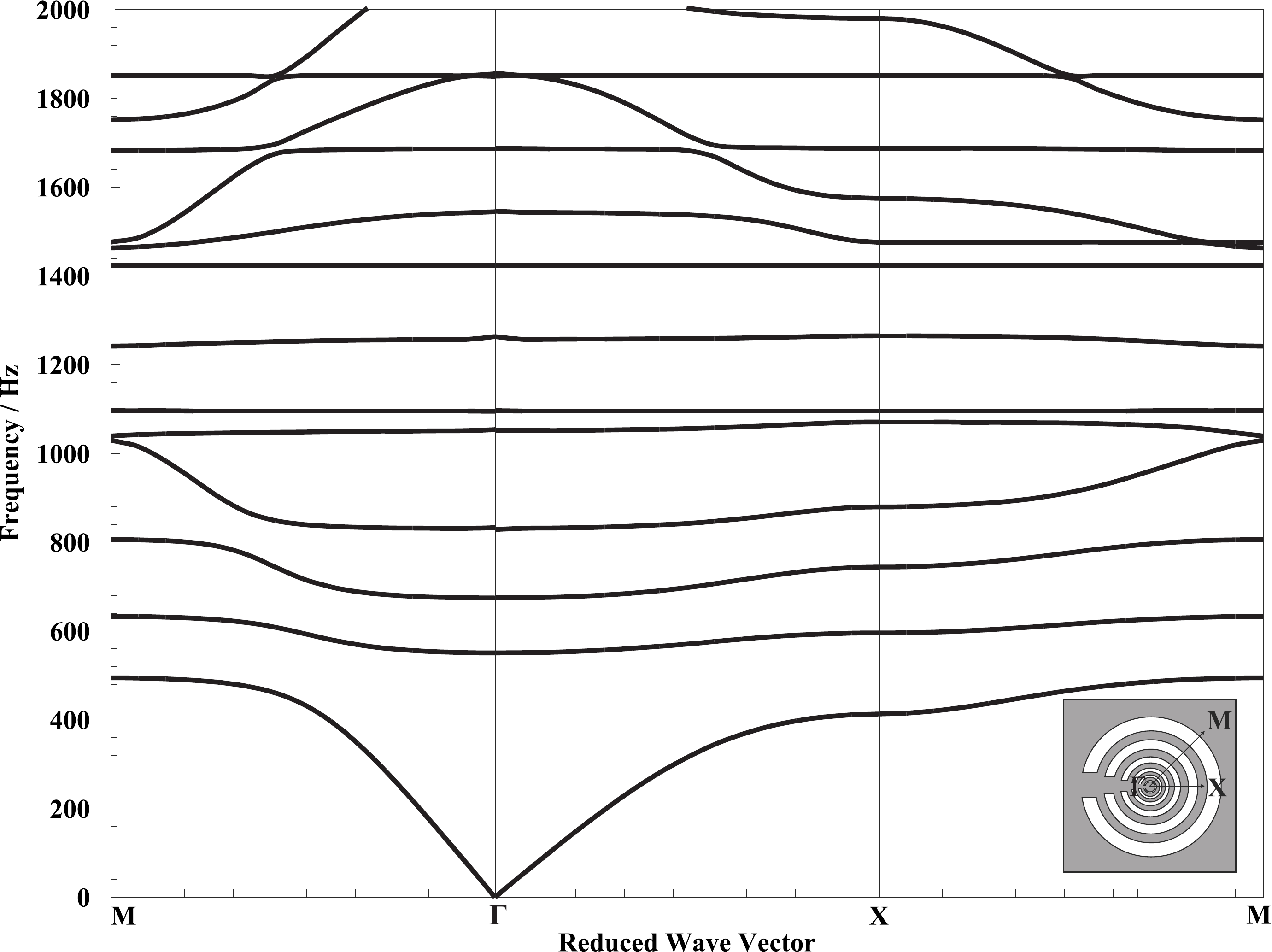}
			\vskip -0.5mm \caption{Finite Element computed band structure for a six concentric Matryoshka system.}
\label{6matband}
	\end{figure}

Applying periodic boundary conditions, to replicate an infinite array of these Matryoshka inclusions in a square array with lattice parameter $a$ = 15.5 mm, the acoustic band structure can be computed. The dimensions of each C-shaped resonator are designed so that they can be placed concentrically inside each other. The largest C-shaped resonator has an, external diameter = 22.5 mm, and an internal diameter = 14.1 mm with a slot width 11.3 mm. Meanwhile the smallest of the nested resonators has an external diameter = 13.2 mm, and an internal diameter = 10.9 mm with a slot width 3.1 mm. Figure \ref{6matband} presents the Finite Element computed band structure in all high symmetry directions.

The band structure has been computed for the first ten Eigenvalues, by varying the wave vector in the first Brillouin zone. It can be seen that a Matryoshka system, with many individual resonating units, induces the formation of multiple band gaps. Due to the periodic nature of these inclusions, this sonic crystal system possesses the characteristic Bragg band gaps, although it is hard to identify which bands are attributed to the separate band gap formation mechanisms. A conventional sonic crystal system with a lattice parameter $a$ = 15.5 mm should possess a Bragg band gap around 1120 -- 1360 Hz, therefore the other band gaps present in the band structure must be caused by the acoustic resonance of each C-shaped inclusion. It is can be seen that the induced resonance band gaps are complete acoustic band gaps, inhibiting wave propagation across all lattice planes, without the need for a large packing fraction as found with the characteristic Bragg band gap.
\begin{figure}[!htp]
	\centering
		\includegraphics[trim = 0mm 0mm 0mm 0mm,clip,angle=0,width=0.85\columnwidth]{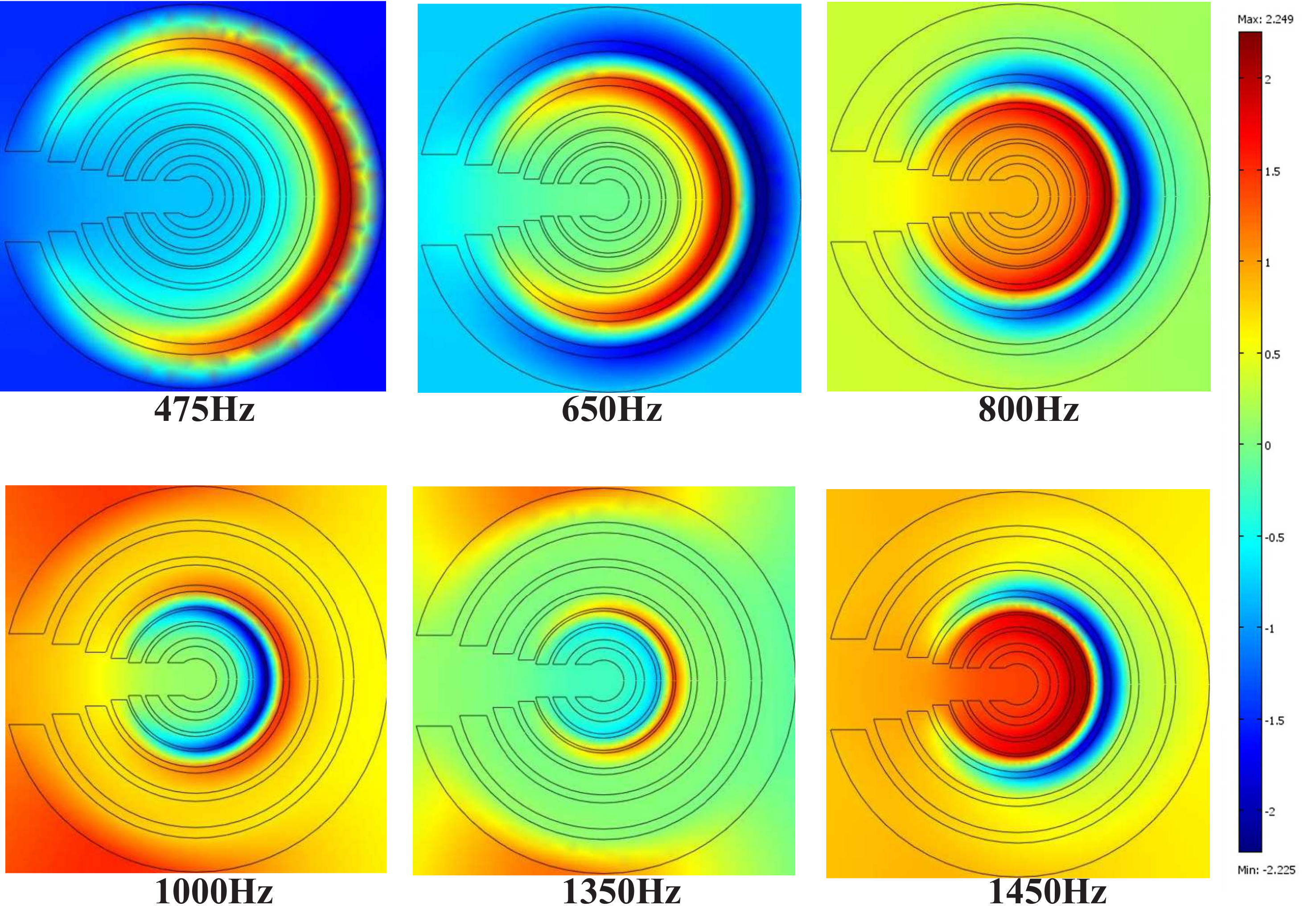}
		\	\vskip -0.5mm \caption{(Colour online) Finite Element computed pressure distribution inside the six concentric Matryoshka system.}
\label{6matpress}
	\end{figure}
For completeness, Finite Element Methods have been employed to obtain a transmission spectrum for this array. A 10 $\times$ 10 array of the Matryoshka inclusions (each containing six concentric C-shaped resonators) is described in Comsol. The spectrum extends to 2000 Hz, and demonstrates the appearance of multiple regions of attenuation, owing to the individual resonances of the six C-shaped resonators as well as a Bragg band gap. The first attenuation band is caused by the individual resonance of the largest diameter resonating inclusion, spanning 400 -- 600 Hz. Five more regions of attenuation can be seen spanning, 600 -- 740 Hz, 740 -- 880 Hz, 880 --  1120 Hz, 1120 Hz -- 1360 Hz and 1360 -- 1500 Hz. 
	
Figure \ref{6matpress} presents the corresponding pressure diagrams computed in the region of each band gap present in the frequency spectrum. It can be seen that each individual resonator experiences an increase in pressure inside the cavity, caused by the acoustic resonance of each C-shaped inclusion. This allows us to confirm the band gap formation mechanism that is responsible for each region of attenuation present in the frequency spectrum.
\begin{figure}[!htp]
	\centering
		\includegraphics[trim = 0mm 0mm 0mm 0mm,clip,angle=0,width=0.85\columnwidth]{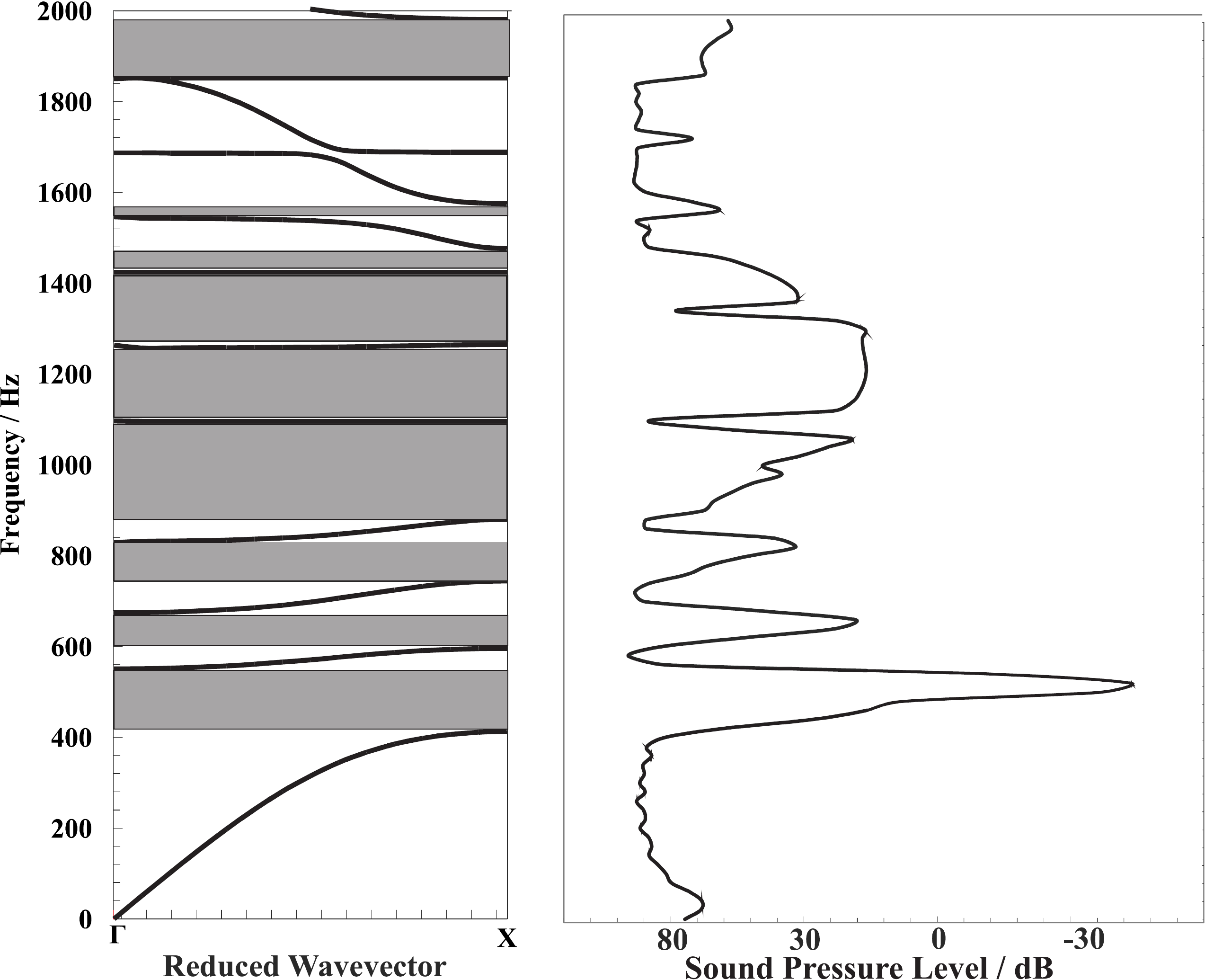}
			\vskip -0.5mm \caption{A comparison of the Finite Element computed band structure with the Finite Element computed frequency spectrum for a six concentric Matryoshka system.}
\label{6matcomp}
	\end{figure}

For comparison, Figure \ref{6matcomp} shows both the Finite Element computed band structure, limited to the $\Gamma X$ direction, and the computed frequency spectrum. The frequencies at which the band gaps occur in the band structure are in good agreement with the regions of attenuation present in the transmission spectrum. A small attenuation band is present in the transmission spectrum at around 1700 Hz. At the corresponding frequency in the band structure, an anticrossing region appears, induced by the level repulsion effect. Since the resonances are very close in frequency to the frequency that satisfies the Bragg condition, the two band gap regimes appear to overlap in this Matryoshka sonic crystal. Resonance scattering occurring in the same frequency range as Bragg scattering favours the formation of broad band gaps.

\section{Conclusion}
The proposed Matryoshka sonic crystal offers a viable solution to overcome the inherent dependence on spacing experienced with conventional sonic crystal designs. It has been discovered that such systems can form multiple resonance band gaps in the lower frequency region, below that of Bragg formation. These resonance bands can be combined to form broad regions of attenuation, either by selecting close acoustic resonances or further by tuning the structure to combine the characteristic Bragg band gap with the resonance band gaps.

The proposed six shell Matryoshka design is particularly suited for noise barrier applications. Although road traffic noise is essentially broad band in nature (due to a large number of very different vehicles that move at different velocities), it often has a defined maximum frequency. A study by Sandberg \cite{Sandberg} found that a multi-coincidence peak in the tyre-noise spectra is observed around 1000 Hz. The simulation results obtained using a six shell concentric Matryoshka system demonstrates the active frequency range spans 400 -- 1600 Hz, providing decent levels of attenuation across this range. 

Moreover, the experimental results provided for the single C-shaped locally resonant sonic crystal \cite{chalmers}, offer decent levels of attenuation for application as a noise attenuation solution.


\begin{thebibliography}{90}

\bibitem{Perez: Sound}
J. V. Sanchez-Perez \emph{et al.}, Sound Attenuation by a Two-Dimensional Array of Rigid Cylinders, Phys. Rev. Lett., \textbf{80}, 5325 (1998).

\bibitem{Sanchez:Acoustic}
J. V. Sanchez-Perez \emph{et al.}, Acoustic Barriers based on Periodic Arrays of Scatterers, Appl. Phys. Lett., \textbf{81}, 5240 (2002).

\bibitem{Liu: Locally}
Z. Liu, Locally Resonant Sonic Materials, Science, \textbf{289}, 1734 (2000).

\bibitem{chalmers}
L. Chalmers, D. Elford \emph{et al.}, Acoustic Band Gap Formation in Two-Dimensional Locally Resonant Sonic Crystals Comprised of Helmholtz Resonators, International Journal of Modern Physics B, \textbf{23}, 20-21, 4234 (2009). Proceedings of the 32nd International Workshop on Condensed Matter Theories, Loughborough, August 2008, World Scientific Publishing, CMT, \textbf{24}, 302 (2010).

\bibitem{Hu: Two}
X. Hu, C. T. Chan \& J. Zi, Two-Dimensional Sonic Crystals with Helmholtz Resonators, Phys. Rev. E., \textbf{71}, 055601 (2005).

\bibitem{movchan}
A. B. Movchan and S. Guenneau, Split-ring Resonators and Localized Modes, Phys. Rev. B., \textbf{70}, 125116 (2004).

\bibitem{Wu}
L. Y. Wu and L. W. Chen, Wave propagation in a 2D sonic crystal with a Helmholtz resonant defect, J. Phys. D: Appl. Phys., \textbf{43}, 055401 (2010).
 
\bibitem{Comsol}
Comsol Inc, Comsol Multiphysics v3.5a., Sweden (2007).

\bibitem{wu:level}
T. T. Wu and Z. G. Huang, Level Repulsions of Bulk Acoustic Waves in Composite Materials, Phys. Rev. B., \textbf{70}, 214304 (2004).

\bibitem{Sainidou}
R. Sainidou and N. Stefanou, A. Modinos, Formation of Absolute Frequency Gaps in Three-Dimensional Solid sonic Crystals, Phys. Rev. B., \textbf{66}, 212301 (2002).

\bibitem{harrison}
W. A. Harrison, Solid State Theory, Dover (1980).

\bibitem{Sandberg}
U. Sandberg, The Multi-coincidence Peak around 1000 Hz in Tyre/Road Noise Spectra, Proceedings of Euronoise, Naples, Italy (2005).

\end{thebibliography}
\end{document}